\documentclass[floatfix,aps,prb,a4paper,showkeys,reprint,superscriptaddress,nobibnotes]{revtex4-1}
\usepackage{amssymb,amsmath,mathrsfs}
\usepackage{graphicx}
\usepackage{epstopdf,bm,natmove}
\usepackage{bm,graphicx}
\usepackage[bookmarks, % add hyperlinks
plainpages=false]{hyperref}
\hypersetup{ %typesetting for online version
    colorlinks,%
}
\usepackage[usenames,dvipsnames]{color}
\usepackage{soul}
\newcommand{\sub}[1]{_{\mathrm{#1}}}
\newcommand{\unit}[1]{\,\,{\mathrm{#1}}}
\newcommand{\amolf}{Center for Nanophotonics, FOM Institute AMOLF, Science Park 104, 1098 XG Amsterdam, The Netherlands}
\newcommand{\dtu}{DTU Fotonik, Department of Photonics Engineering, \O stedsplads 343, DK-2800, Denmark}
\newcommand{\ornstein}{Condensed Matter and Interfaces, Debye Institute for Nanomaterials Science, Princetonplein 1, 3584 CC Utrecht, The Netherlands}

\begin{document}
\author{Per Lunnemann}
\email{plha@fotonik.dtu.dk}
\affiliation{\amolf}
\affiliation{\dtu}
\author{Freddy T. Rabouw}
\author{Relinde J. A. van Dijk-Moes}
\author{Francesca Pietra}
\author{Dani\"{e}l Vanmaekelbergh}
\affiliation{\ornstein}
\author{A. Femius Koenderink}
\affiliation{\amolf}

%\date{\today}

\title{Calibrating and Controlling the Quantum Efficiency Distribution of Inhomogeneously Broadened Quantum Rods Using a Mirror Ball}	

\begin{abstract}
We demonstrate that a simple silver coated ball lens can be used to accurately  measure the entire distribution of radiative transition rates of quantum dot  nanocrystals. This simple and cost-effective implementation of Drexhage's method that uses nanometer-controlled optical mode density variations near a mirror, not only allows to extract calibrated ensemble-averaged rates, but for the first time also to quantify the full inhomogeneous dispersion of  radiative and non radiative decay rates across thousands of nanocrystals. We apply the technique to novel ultra-stable CdSe/CdS dot-in-rod emitters. The emitters are  of large current interest due to their improved stability and reduced blinking. We retrieve a room-temperature  ensemble average quantum efficiency of $0.87\pm0.08$ at a mean lifetime around 20 ns. We confirm a log-normal distribution of decay rates as often assumed in literature and we show that the rate distribution-width, that amounts to about 30\% of the mean decay rate, is strongly dependent on the local density of optical states. 
\end{abstract}
\maketitle
Over the past decades, the development of bright  single  emitters across the visible and near infrared spectrum has experienced major progress. Today there exist numerous different type of single emitters such as organic dye molecules, chemically synthesized II-VI nano crystals, and epitaxially grown III-V semiconductor quantum dots. Their usage spans from fluorescence microscopy,\cite{Resch-Genger2008,Lichtman2005} light emitting diodes, \cite{Caruge2008} lasers, optical amplifiers,\cite{VanderPoel2006,Eliseev2001} and  quantum photonics.\cite{Malik2010,Hennessy2007a}
Especially the semiconductor based emitters are promising due to the possibility of tailoring their emission wavelength through size, their exceptionally large oscillator strength, and the potential of integration with photonic structures using top down semiconductor fabrication, or bottom up self assembly techniques. 
A common goal is to optimize light emission and extraction and to reduce undesirable non-radiative decay processes. This may either be done through engineering of the electronic structure of the emitter through a suitable choice of materials and synthesis routes, or by tailoring the photonic environment.\cite{Lund-Hansen2008,Julsgaard2008,Frimmer2011,Frimmer2012} The radiative emission rate depends on the local density of optical states (LDOS), as given by Fermi's golden rule\cite{Sprik1996,Loudon2000}, and may be controlled by structuring the surrounding material. Photonic crystals as well as plasmonic optical antennas have been successfully demonstrated to enhance and guide light emission.\cite{Lodahl2004,Lund-Hansen2008,Julsgaard2008,Kuhn2006,Anger2006} To assess the success of strategies to improve emitters, whether through chemistry or photonics, it is essential to have a method that determines radiative and nonradiative decay rates, as well as quantum efficiencies in a rapid, yet calibrated manner. 

An established technique to measure the intrinsic non-radiative and radiative time constants of an emitter, is based on applying a well defined change of the LDOS by changing the distance of an emitter to a  planar mirror, as pioneered by Drexhage \emph{et al}\cite{Drexhage1966}. By measuring  the fluorescence decay rate, which is the sum of non-radiative and radiative decay constants, as a function of LDOS, it is possible to extract the contribution of non-radiative decay, since only the radiative decay rate varies with LDOS. As opposed to brightness comparisons or integrating sphere measurements, Drexhage's method is absolute, requires no reference, and is artifact free. The method has been applied to rare earth ions,\cite{Drexhage1966,Snoeks1995,Amos1997} organic dye molecules,\cite{Kwadrin2012,Danz2002} epitaxially grown  III-V semiconductor QDs, \cite{Johansen2008,Stobbe2009,Andersen2010} and chemically synthesized II-VI semiconductor QDs. \cite{Leistikow2009,Kwadrin2012,Brokmann2004} Unfortunately, these experiments are cumbersome to implement\cite{Leistikow2009}, requiring either elaborate deposition techniques or sample-specific etching methodologies to vary the emitter-mirror distance\cite{Johansen2008,Stobbe2009,Andersen2010}.  One of the simplest reported implementations of the technique is based on a gray-tone lithography  to fabricate an inclined spacer layer on top of the emitters\cite{Kwadrin2012}. While the fabrication  is relatively simple, a required UV exposure may photobleach emitters already prior to testing. Furthermore, UV lithography  materials tend to  show strong background fluorescence and involve solvents that put the integrity of the emitters at risk. Finally, we note that the method is not suitable for calibrating an actual device. \emph{I.e.} it requires a sample dedicated for calibration.
Another simple implementation{\cite{Brokmann2004}} is based on measuring the decay rate of emitters at two positions, one in front of an PMMA-air interface and one with no interface by simply adding a droplet of index matched PDMS on top of the PMMA. While the method is simple, the controlled change of the LDOS is very small ($\sim 15\%$), thus limiting the method to emitters with a large quantum efficiency. Secondly, the simplicity of only measuring at two distances comes at the cost of compromising the robustness of the calibration (fitting a straight line to two points).
Recently micro-mechanical techniques to vary emitter-mirror distance were introduced, which avoid such chemistry-related issues, and are excellent for single-molecule studies. However, micromanipulation is technically challenging and not easily scaled to obtain ensemble  statistics beyond a few tens of molecules\cite{Buchler2005,Frimmer2012a,Chizhik2011}. 

Here, we report on an implementation of Drexhage's method that has two main benefits: 
First, the method serves as simplification of a well-known measurement technique to calibrate the ensemble radiate decay rate.  Without the need for lithography or sample-specific chemistry, the method allow for significant savings in terms of measurement efforts, without compromising the fidelity. Second, and more importantly, the method allows retrieving information beyond ensemble-average rates. We demonstrate this benefit by showing how the entire distribution of decay rates of a huge ensemble of $10^{5}$ quantum dots depends on the LDOS. We apply the method on a novel promising CdSe/CdS rod structure that shows a reduction of the universal phenomenon of fluorescence intermittency, \emph{i.e.} blinking. As with all currently available quantum dots, the ensembles are very inhomogeneous, but using this method we demonstrate, for the first time, exactly how the entire ensemble of rates are distributed and modified by the LDOS using the massive acquired data set.

\begin{figure}
\includegraphics[width=8.2cm]{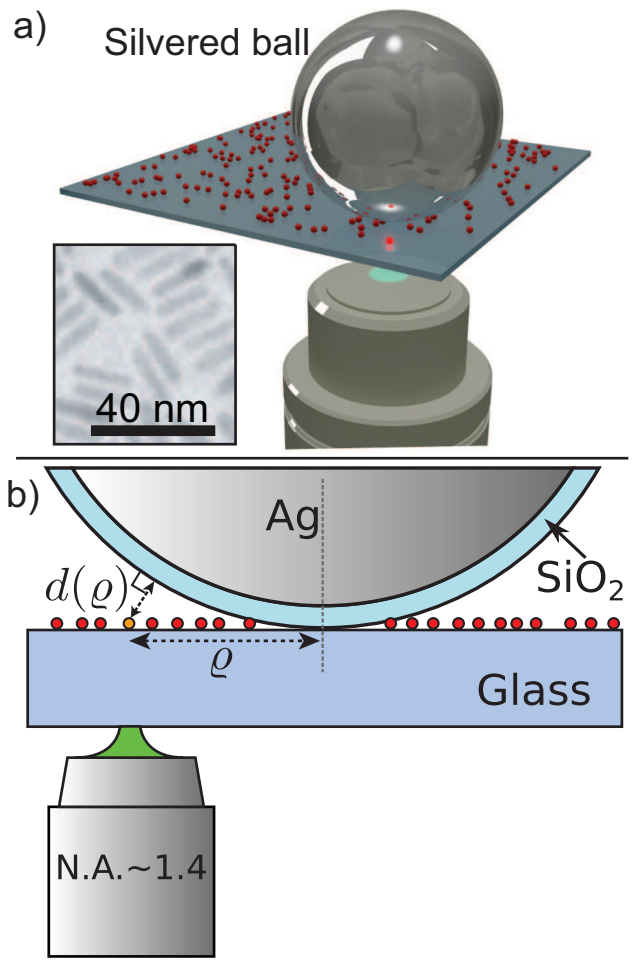}
  \caption{ a) Illustration (not to scale) of the mirror configuration. The silver coated ball lenses are mounted in a tripod configuration for easier handling and fixation of the position. Red spheres indicates the nanorods while green is the pump light. The inset shows a TEM image of the nanorods
  b) Cross sectional illustration of the measurement setup (not to scale).
 \label{fig:setup}}
\end{figure}

The essence of our Drexhage method implementation is that we realize  a precise yet low cost mirror that can be used to perform the entire measurement on a single sub-millimeter size sample,   and  that may be reused on many samples. The mirror is created by evaporating a thin layer of silver onto a commercially available ball-lens. The spherical mirror is then put on top of a thin glass substrate that is spin coated with emitters, see figure
 \ref{fig:setup}. 

In the following section, we first show how rates for an entire ensemble of $10^5$ emitters are extracted. Subsequently, we uncover that the huge distribution of rates is not associated at all with a spread of nonradiative decay, but rather a spread in intrinsic radiative decay rates in combination with a dipole-orientation dependence caused by the substrate.

\section{Results and discussion}
\begin{figure}
\includegraphics[width=8.46cm]{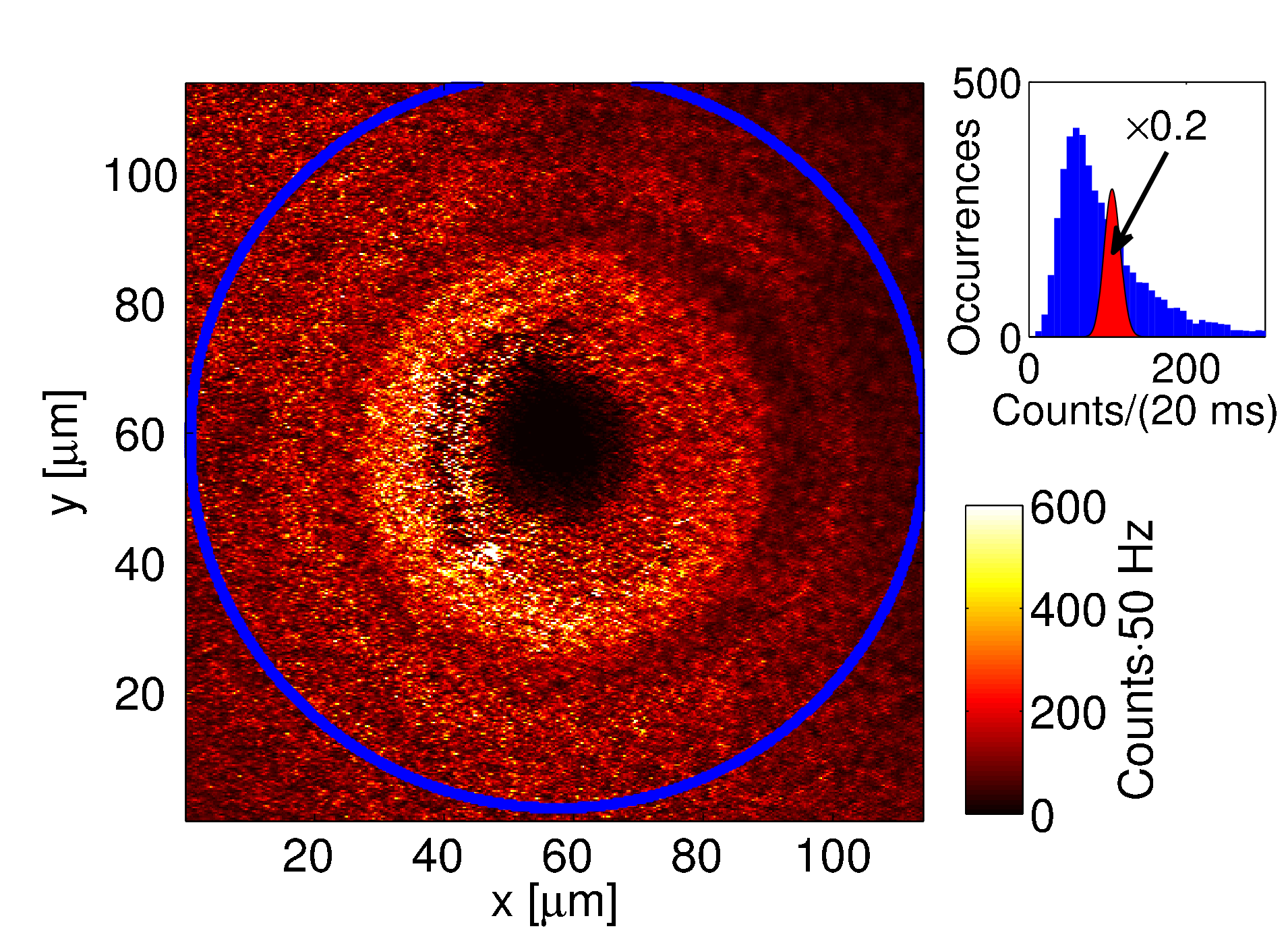}
\caption{Detected counts of a 400 by 400 step confocal scan corresponding to $\sim 115\times 115\unit{\mu m}$ scan.  The blue ring indicates points in a concentric ring around the spherical mirror center that are histogrammed (top right, blue). On top, the corresponding Poisson distribution, scaled by 0.2 is plotted (red). The mismatch between the distributions signify that the granularity of the image is associated with the discrete nature of the emitters rather than Poisson photon counting noise.\label{fig:confScan}}
\end{figure}

 Figure \ref{fig:confScan} presents a map of the collected photon counts of a confocal scan performed on a $400\times 400$ pixel grid covering a $115\unit{\mu m}\times 115 \unit{\mu m}$ square corresponding to a step size of $290\unit{nm}$. 
We observe concentric rings with a low count region in the middle that is identified as the center  of the sphere. We particularly tuned the nanorod dilution such that at the given collection area and step size we  obtain information beyond  the ensemble averaged rate. 
An example of a histogram of counts from points in a concentric ring about the mirror center is presented in the inset of figure {\ref{fig:confScan}}, together with the corresponding Poisson distribution. The clear mismatch between the two distributions clarifies that indeed the granularity of the image is not Poisson photon counting noise, as would be expected from a homogeneous layer at this count rate. To estimate an upper bound to the mean number of quantum dots probed per pixel, assume that each NR yields the same intensity without photon counting noise. Under this assumption, the entire distribution in counts comes from the shot noise in the number $N$ of NRs per pixel. Taking the ratio of width to mean of the count distribution as estimate for $1/\sqrt{ \langle N\rangle}$, we find, for the points indicated in blue in figure {\ref{fig:confScan}}, that the mean $\langle N\rangle$ is at most 1.2 NRs/px. This is an upper bound since the estimate ignores other evident sources of noise in the histogram. From this we conclude that  granularity is associated with the discrete nature of the emitters.
While in each pixel, we probe the decay rate of approximately one or few NRs, taking pixels in concentric rings around the spherical mirror center together will add to an ensemble of many quantum dot-in-rod emitters at a fixed emitter-mirror separation distance (see figure
\ref{fig:confScan}). In the central low count region, all NRs are in close vicinity to the metal surface. Radiative emission is therefore quenched due to resonant coupling to the surface plasmons polariton (SPP) mode.\cite{Amos1997} Moving radially out from the center, the observed fringes in intensity are largely due to  the standing wave that the pump laser and its reflection form at the interface. The  thin streak of lower counts within the left part of the first ring we attribute to a surface irregularity  such as a minute scratch, on the mirror surface.

\begin{figure}
\includegraphics[width=8.5cm]{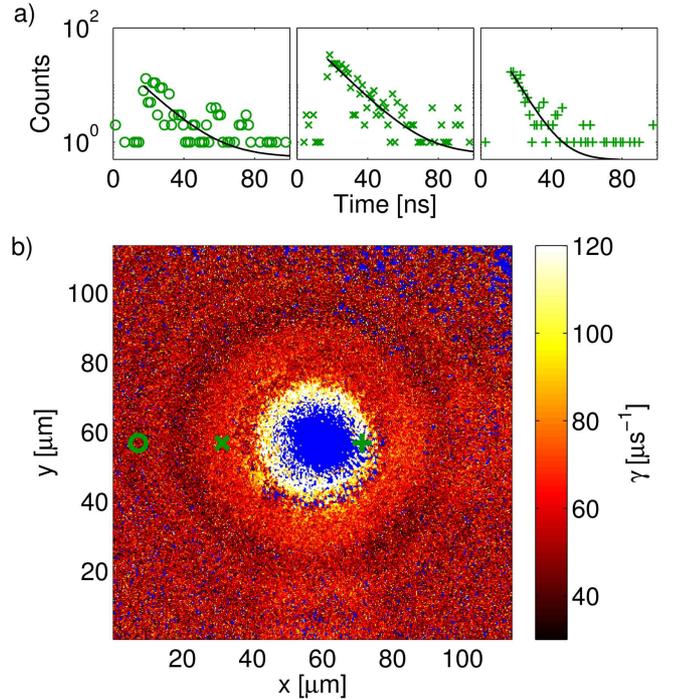}
\caption{a) Example  measured decay traces at three isolated positions, each collected over 20 ms. A single exponential fit is shown as a black solid line. b) The map of extracted decay rates \emph{versus} lateral position under the mirror shows a clear modulation. Blue pixels  mark positions with counts below 40 where no fit was attempted. The three positions associated with a) are indicated with a green circle (a) left), a cross (a) middle) and a plus (a) right).\label{fig:decayrateMap}}
\end{figure}

For each position in figure \ref{fig:confScan}, we histogram the arrival times of the detected photons relative to the laser pulse in bins of 1.32 ns (an 8-fold coarsening relative to the timing card resolution). From the accumulated histogram, we extract a  fluorescence decay rate  by fitting a single-exponential decay with an added background. Importantly, since we are counting the photons arriving within specified time bins, our data is characterized by having a Poissonian probability distribution within each time bin. Using the maximum-likelihood fitting procedure,{\cite{Bajzer1991}} Poissonian statistics implies minimizing a merit function of the form $-\sum_{i=1}^{N}\{ D(t_{i})\log\left[F_{\gamma}(t_{i})\right]-F_{\gamma}(t_{i}) \}$, where $D(t_{i})$ is the measured counts in the $i$th bin and $F_{\gamma}$ is the fit function with fitparameter(s) $\gamma$. While this method is common practice in the field of time correlated single photon counting, we note that the often used least squared residual merit function applies for Gaussian statistics. Although, for large counts, the Poissonian distribution approaches the Gaussian distribution, the correct choice of a merit function implied by Poissonian statistics is crucial for our experiment with low counts.
Examples of measured decay traces and the fitted single exponential curve for three isolated measurement positions are presented in figure \ref{fig:decayrateMap}a). The location of these three  pixels is indicated in the complete map  of  extracted decay rates in figure \ref{fig:decayrateMap}b).
We find a clear modulation of the decay rate in concentric rings around the mirror center,  with the largest decay rates at positions in close vicinity to the mirror surface.  Blue pixels in the map mark  positions with  below 40 counts in the entire decay trace, where no fit was attempted. To confirm that using a single exponential function applies to the current emitters, we carried out several measurements on single NRs far away from the mirror, integrating over 150 s, see supporting material available online. The acquired histogram from the total set of data exhibited excellently single exponential behavior with an estimated decay rate of $56.36\unit{\mu s^{-1}}$ and an associated standard deviation of only $\pm0.05\%$. While, the confidence intervals of the extracted decay rate at each pixel in figure {\ref{fig:decayrateMap}} differ significantly primarily due to the span in counts ranging from 40 counts to several hundreds, we note that from the long integration time experiment on single NRs, the standard deviation of the estimated rates of each 20 ms time frame was $\sim\pm14\%$ for each $20\unit{ms}$ (see supplementary material), thus confirming that indeed it is possible to establish decay rate dynamics with counts on the order of 100.\cite{Galland2011}
	
\begin{figure}
\includegraphics[width=8.5cm]{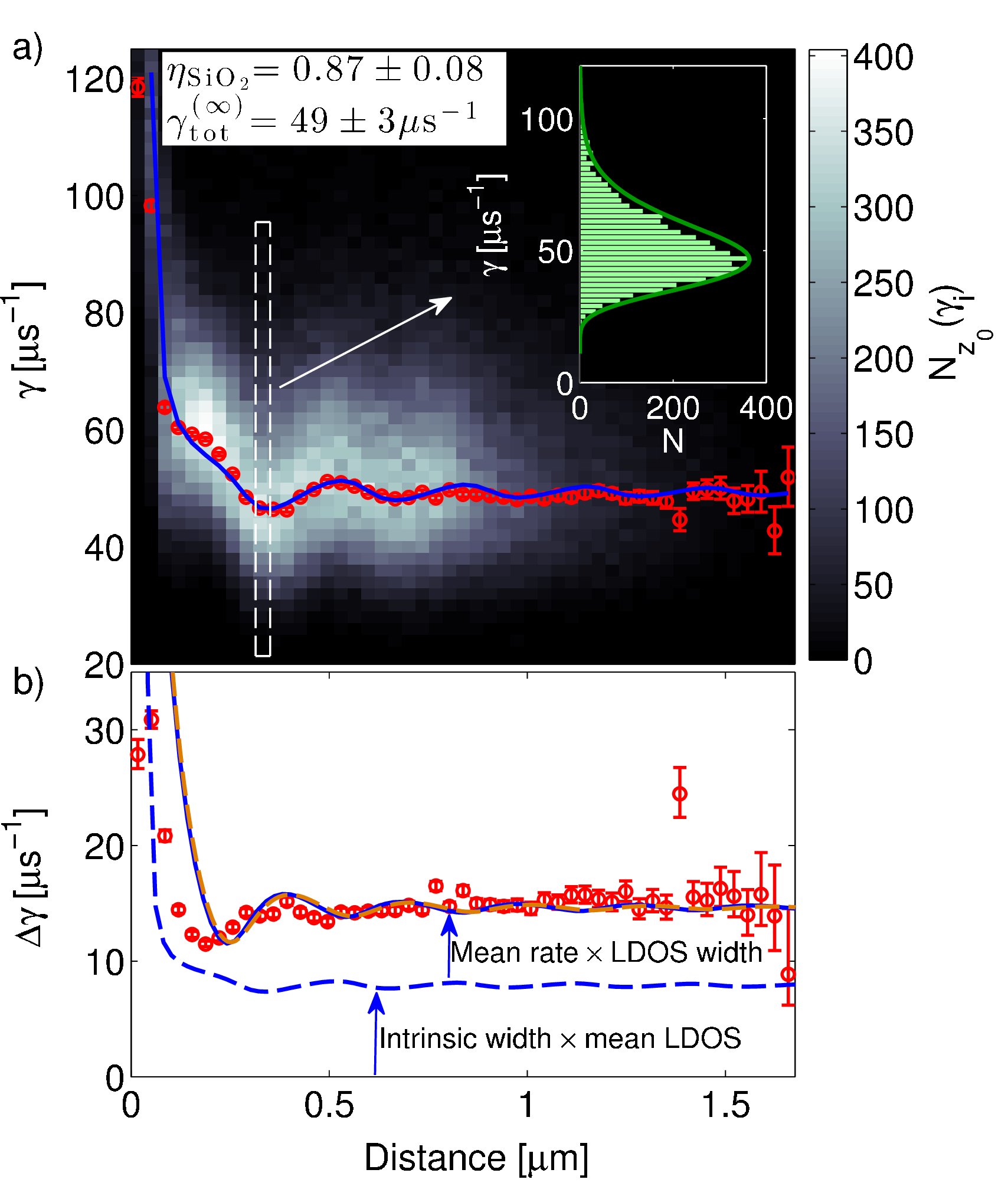}
  \caption{a) Measured decay rates as a function of distance to the mirror. Plotted in gray colors is the histogram of extracted decay rates for a given distance. The inset shows an example of the acquired histogram of decay rates at a distance $0.32\unit{\mu m}$, indicated by the dashed box. Red markers indicate the extracted most frequent, $\gamma_0$ for each distance based on a fitted normal distribution. The error bars indicate the 95$\%$ confidence interval of the extracted decay rate $\gamma_0$. The blue curve is a weighted fit based on equation \ref{eq:fitrate} assuming a central wavelength of 610 nm with fitted quantum efficiency $\eta$ and intrinsic total decay rate, $\gamma\sub{tot}^{(\infty)}$ stated in the white box. b) Red data points with error bars: the extracted standard deviation, $\Delta\gamma$, of the lognormal distribution of decay rates.  Dashed line:  expected histogram width assuming an intrinsic radiative rate distribution width of  $\Delta \gamma\sub{rad}^{(\infty)}=7.5\unit{\mu s^{-1}}$, if all emitters would experience the same, mean LDOS.  Solid blue line: effect of orientation-dependent LDOS, assuming the same mean decay rate for all NRs,  added to the dashed line. Orange dashed curve:  as solid blue line, but also including the wavelength-spread induced LDOS variation.  The wavelength inhomogeneity is unimportant in determining the LDOS dependence of the decay rate distribution width.\label{fig:drexhage}}
\end{figure}

To quantitatively extract radiative and nonradiative rates, we convert the pixel coordinates in the 2D  map to emitter-mirror separation, so that the data  can be compared to LDOS calculations. We  identify the contact point, $\bf{\varrho_0}$, of the sphere with the substrate as the center of the rings in figure \ref{fig:confScan} with an estimated accuracy of 5 nm. Next, as a measure for the emitter-mirror separation  we calculate the radial distance $d$ from the mirror surface to an emitter on the substrate as
$d(\bm{\varrho})=\sqrt{R^2+|\bm{\varrho}-\bm{\varrho}_0|^2}-R$, where $R$ denotes the radius of the coated ball lens. We bin all measurement points into a set of concentric bands $d_i$, defined by having emitter-mirror separation   $d_i-\delta d/2 \leq d(\varrho) <d_i+\delta d/2$ . For each band, we calculate a  histogram of the extracted decay rates, using a bin size of $2\unit{\mu s^{-1}}$. The extracted decay rate histograms are plotted in figure \ref{fig:drexhage}  as a function of distance, using a  $\delta d=34\unit{nm}$. Each histogram typically contains decay rates fitted to approximately  3000 pixels.
We observe a wide spread in decay rates  for each distance, with a mean decay rate of around 50 $\mu$s$^{-1}$ (decay time 20 ns), and a relative distribution width of about 50\%. Since the statistical uncertainty associated with the fitted decay rate of each pixel  (typically $\sim 3\unit{\mu s^{-1}}$, comparable to a single histogram bin width) is much smaller than the observed spread of decay rates, we attribute the width   to inhomogeneous broadening of the NRs.  Notably, the entire histogram clearly shifts depending on distance to the mirror. To define the mean decay rate and the width of the distribution, the histogram at each distance is  fitted with a lognormal-distribution defined through
\begin{equation}
P(\gamma;\mu,\sigma) = \frac{1}{\gamma \sigma\sqrt{\pi}}\, e^{-\frac{(\ln\gamma - \mu)^2}{\sigma^2}},
\end{equation}
 where $\sigma$ and $\mu$ are the standard deviation and mean of $\ln \gamma$, respectively. The most frequent decay rate, $\gamma_0$, and the standard deviation, $\Delta\gamma$, are related to $\mu$, and $\sigma$ as
 \begin{align}
\gamma_0&=e^{\mu-\sigma^2}\\
\Delta\gamma&=\sqrt{\left(e^{\sigma^2}-1\right)e^{2\mu+\sigma^2}}
\end{align}
As seen from the inset in figure \ref{fig:drexhage} , the log-normal distribution describes our data excellently. The fitted most frequent decay rate, $\gamma_0$, is plotted with red circles in figure \ref{fig:drexhage}. The error bars indicate a $95\%$ confidence interval of $\gamma_0$. We clearly resolve oscillations of the mean decay rate as well as  the decay rate divergence at very small distances, as expected from the modulation of the  LDOS at a metal interface.  In the following we first discuss the dependence of the mean decay rate on distance to the mirror, and then further discuss the histogram width. 

As in usual ensemble measurements, we combine calculated LDOS and  the measured mean decay rate to fit  values for the intrinsic ensemble average $\gamma\sub{rad}$, and non-radiative decay rate $\gamma\sub{nr}$. Using $\rho(d)$ to denote the calculated LDOS enhancement relative to the LDOS in SiO$_2$ in absence of the mirror,  we fit the measured most frequent  decay rate $\gamma_0 (d)$ to 
\begin{equation}
\gamma_0(d)=\gamma^{(\infty)}\sub{tot}\left\{ 1+\eta\left[\rho(d)-1\right]\right\}. \label{eq:fitrate}
\end{equation}

The two fit parameters represent the ensemble-average total decay rate $\gamma^{(\infty)}\sub{tot}$ for the quantum dot-in-rod emitters in glass, in absence of the mirror,  and  the ensemble-average quantum efficiency $\eta$ (again in absence of the mirror).   The calculation of  $\rho(d)$ assumes a stratified structure consisting of a semi-infinite glass substrate, vacuum, $\mathrm{SiO}_2$, and a semi-infinite Ag slab. The assumption of a stratified parallel layered structure is well approximated from the fact that the maximum angle between the tangent of the mirror and the glass interface is $1.4^{\circ}$.  The calculations uses the established integration methodology reported in \cite{Paulus2000} and the parameters listed in table \ref{tab:params}. The refractive indices of the $\mathrm{SiO}_2$ and Ag layer were measured by ellipsometry. As wavelength we take the center emission wavelength from the measured ensemble emission spectrum,  and take the emitter location as $20\unit{nm}$ above the glass substrate to account for the SiO$_2$ shell.  
\begin{table}
\caption{\label{tab:params} Used parameters for calculating the LDOS. Parameters marked by * indicate a measured quantity, while those without are estimates.}
\begin{tabular}{lr}
\textbf{Parameter}  & \textbf{Value}  \\ 
\hline  *Thickness $\mathrm{SiO_2}$& $35\unit{nm}$  \\ 
  *Refractive index $\mathrm{SiO_2}$& $1.523$  \\
  *Refractive index Ag & $0.0751 + i4.191$\\
  Height above glass substrate of emitter & $20\unit{nm}$\\
  * Centre emission wavelength & $610\unit{nm}$\\
\end{tabular} 
\end{table}

We find an excellent fit to the mean decay rate when we assume isotropically oriented transition dipole moments, in good agreement with earlier results on CdS/ZnS quantum dots\cite{Leistikow2009}. In this work, Leistikow \emph{et al}. simulated decay traces assuming a 2D degenerate dipole moment in each quantum dot in an isotropic ensemble, exactly in a Drexhage geometry. In spite of the degeneracy, the conclusion is that rates fitted to the decay traces should fit very well with the isotropically averaged LDOS. This is opposed to self-assembled III-V semiconductor quantum dots that are strongly oriented parallel to the plane of growth\cite{Johansen2008}. We get a fitted ensemble-average total decay rate of $\gamma^{(\infty)}\sub{tot}=49\pm3~\mu$s$^{-1}$ and an intrinsic ensemble-average quantum efficiency of $\eta = 0.87\pm 0.08$. Previous experimental work on CdSe/CdS NRs\cite{Carbone2007,Talapin2003} concluded on basis of strongly polarized emission that the emission dipole of NRs is oriented along the long axis of the NRs.  However, owing to the thick SiO$_2$ shell, the NRs in our experiment do not necessarily lie flat on the surface, explaining that the best fit is obtained assuming  isotropic dipole orientation.  The fitted values show that at a mean total fluorescence lifetime of 19 ns, and quantum efficiency of around 90\%, the CdSe dot in CdS rod system is highly promising for applications as a bright emitter. The fitted quantum efficiency  is well above the ensemble quantum efficiency that we obtain from absorption/emission brightness measurements, in good agreement with previous ensemble studies of quantum dots (QDs) \cite{Kwadrin2012,Leistikow2009}, because dark quantum dots are counted in  ensemble absorption, but not in emission lifetime measurements. A subtle point here is that these particular quantum dot-in-rods in fact do not have a completely dark state, but rather show intermittent switching between a bright state, and a gray state that is only approximately three times dimmer than the bright state. As the nanorods in the gray state have an estimated quantum efficiency around four times lower than the bright state, and do contribute to the decay rate traces, the fitted $\eta=0.87\pm0.08$ provides an underestimate for the actual quantum efficiency of the bright state.   

The main advantage of our new method to calibrate fluorophores, in addition to simplicity, is that information \emph{beyond} ensemble average rate and quantum efficiency is obtained, in form of the full histogram of decay rates, comprising thousands of emitters.  We now turn to a  discussion of the histogram beyond the average rate.  In previous reports,  log-normal distributions have been assumed to describe the  decay rate distribution of ensembles of quantum dots.~\cite{VanDriel2007,Nikolaev2007,Kwadrin2012} However, in those reports only a  single decay trace for an entire ensemble of dots was recorded, which was fitted to the decay expected for a log-normal distribution of rates.  This procedure relies  on an assumed distribution of rates, while in fact  the ensemble-average decay transient that is measured might be fitted by many different  non-single exponential fit functions.
We here provide a method that can directly prove or disprove the suitability of a particular rate distribution.  For this particular system, we find that the rates follow a log-normal distribution, cf. inset in figure \ref{fig:drexhage}.   Importantly, when we reverse the order of ensemble averaging, \emph{i.e.} fitting pixel-averaged decay traces with a decay law for log-normally distributed rates, we find a similar most frequent decay rate and distribution width, which provides an \emph{a posteriori} validation of previous approaches.\cite{VanDriel2007,Nikolaev2007,Kwadrin2012}

Considering figure \ref{fig:drexhage}, the width of the decay rate distribution evidently varies with LDOS, \emph{i.e.},  with mirror-emitter separation. At small separations to the mirror, where the mean decay rate depends strongly on the distance to the mirror, we find a strong change in the width of the rate distribution that follows the LDOS variation.  An LDOS dependence of the distribution width is induced through various mechanisms.  To first order, the combined effect of LDOS on the width can be written as
\begin{equation}
\Delta\gamma(z)=\Delta\gamma\sub{nr} + \Delta \gamma\sub{rad}^{(\infty)}\cdot \rho(z) + \gamma\sub{rad}^{(\infty)}\cdot \Delta \rho (z) . \label{eq:width}
\end{equation}
Here, the first two terms describe the effect of an intrinsic inhomogeneous broadening, for instance due to size and shape dispersion, separated in  a LDOS-insensitive nonradiative contribution $\Delta\gamma\sub{nr}$ as well as a LDOS-dependent radiative rate distribution. If  all sources are subject to the same LDOS variation, all the radiative rates within the distribution are multiplied by the LDOS. Therefore, if all sources are highly efficient, the width of the distribution is directly proportional to the LDOS at the emission frequency , so that the width  scales as $\Delta \gamma =\Delta \gamma^\infty\cdot\mathrm{\rho}(z)$. Conversely, the width of the rate distribution would remain independent on distance if the spread is purely due to a distribution in nonradiative decay constants, $\Delta\gamma\sub{nr}$. The third term in equation \ref{eq:width} accounts for a second effect that also affects the distribution, namely that the mirror offers a different LDOS for different transition dipole moment orientations, and emitters with different emission wavelengths in the inhomogeneous ensemble. 
To investigate the role of dipole moment orientations and emitter wavelengths we have calculated the variation $\Delta \rho$ in LDOS for the two scenarios: 1): Taking into account the random dipole orientation, while assuming that all emitters emit with the same wavelength, and 2): Taking into account the measured inhomogeneously broadened ensemble spectrum, while assuming that all emitters experience the orientationally averaged LDOS.  For evaluation of the first scenario, we note that from the calculated LDOS at parallel and perpendicular dipole orientation relative to the substrate normal,  the rate at any dipole orientation is completely known,\cite{Vos2009} and hence the distribution.  We find that  the LDOS distribution induced through inhomogeneous broadening of the emission wavelength is negligible compared to the distribution caused by orientational effects.

In figure \ref{fig:drexhage}b) we plot the different contributions to the measured histogram width,  and compare these with the data.  The intrinsic inhomogeneous decay rate distribution, corresponding to NRs in a homogeneous dielectric environment, contributes approximately half of the observed width.  The other half is contributed by the orientation dependence of the local density of states.  Setting the intrinsic inhomogeneous broadening of the radiative decay rate to a width $\Delta \gamma\sub{rad}^{(\infty)}=7.5\unit{\mu s^{-1}}$  we find a reasonable agreement with the measurements except for distances close to the mirror.
Interestingly, we found the best fit setting $\Delta\gamma\sub{nr}=0$, implying that most broadening is caused by a spread of radiative decay rates, and not nonradiative effects. This  assessment is firstly consistent with the high quantum efficiency fitted to the mean decay rate, which indicates that nonradiative contributions are negligible, and secondly indicates that for the fastest decaying emitters in the ensemble, the quantum efficiency is not necessarily lower than for the slowest ones. With  regard to the contribution of dipole orientation, we note that $\Delta \rho_\theta$ does not vanish far away from the mirror, as a result of the glass substrate causing a dipole-orientation dependence on the LDOS even in absence of the mirror. 
It might be argued that the fact that half of the decay rate distribution width is due to orientation dependence at the glass-air interface, and not to a difference in oscillator strength, is an artifact of the measurement method.  However, we argue that this is by no means a limitation. First, as the glass-air interface is completely understood, it can be corrected for so that the calibration still gives complete access to the entire intrinsic distribution that is otherwise completely inaccessible by any other method. This correction is applicable in any case where the intrinsic distribution is comparable in width to the orientational spread in LDOS. Second, the experiment could easily be modified to use immersion liquid between ball and sample, removing the intrinsic orientation dependence.  Third, and most importantly, if the presence of the glass-air interface is to be deemed an artifact, it must be realized that it is in fact an artifact common to almost all state of the art single quantum dot studies. Almost all state of the art single quantum dot studies aimed at quantifying charge dynamics, Auger recombination, blinking, spectral wandering \emph{etc}. are carried out right at the interface of a glass microscope slide, \emph{e.g.} see refs. {\cite{Galland2012,Galland2011,Carbone2007,Spinicelli2009,Cordones2011}}. The distribution we uncover is hence directly representative for all such studies, and conversely, it is an important realization that all such studies will result in widely distributed values of extracted parameters, unless the glass-air interface and its LDOS is corrected for.

\section{Conclusion}
We have demonstrated a novel implementation of the Drexhage experiment for extracting the ensemble-average radiative decay rate and quantum efficiency of dipole emitters. On one hand, the method serves as a simplification of a well-known measurement technique that allows for momentous savings in measurements efforts. In our method, the measurement procedure is still sequential, as we aim to uncover an entire ensemble statistics. However if just ensemble-average data is desired, using a streak camera one could collect the entire data set in a single shot in a matter of seconds. On the other hand, these huge savings in simplicity open up new and exciting possibilities, \emph{i.e.}, that of acquiring so far inaccessible information of the entire distribution of decay traces from a huge ensemble with $10^{4}\ldots10^{5}$ emitters,  and its dependence on the local density of optical states. The method is based on the use of a very simple metal-coated ball lens  to create a controlled local density of optical states variation. We applied this  method to a novel type of CdSe/CdS dot-in-rod nanocrystals, a novel structure that holds promise to show excellent photostability and the absence of a true dark state, and according to our measurement has a quantum efficiency of 80\% or above.  We anticipate our method to have many applications, since the ball lens is not only easily fabricated, but can simply be placed on top of, and subsequently removed from, any sample substrate. This advantage moves Drexhage calibration from a tedious system specific fabrication procedure to be applied on dedicated test samples, to a method that could even be applied on actual completely functional devices, such as for instance III-nitride light-emitting diodes and organic light-emitting diodes. Furthermore, we have demonstrated that this new technique provides a  unique opportunity to uncover photo-physical parameters beyond ensemble-averaged decay traces. By carefully chosen dilution we operate at just one or a few quantum dot-in-rod emitters per pixel, while we can still collect significant ensemble statistics owing to the large number of pixels for which we obtain signal.  Indeed, we obtain a  full distribution of decay times from single exponential fits to pixels that each represent signal from just one or a few emitters. In previous works,  several authors have advocated that in measurements of larger ensembles the distribution of rates causes the average decay trace to be not single exponential, but best fitted with a log-normal distribution.  In our work we go beyond this indirect evidence as we reverse the order of fitting decay constants, and assembling ensemble data.  For the particular quantum dot-in-rod system we study, rate distributions are indeed log-normal, and we confirm that a log-normal fit to ensemble-averaged decay traces is indeed consistent with the full log-normal distribution of histogrammed rates.\cite{VanDriel2007,Leistikow2009,Kwadrin2012} In the system studied here, the inhomogeneously broadened width of the decay rate distribution  is due in equal parts to  an intrinsic distribution in radiative rate, and additionally dipole-orientation dependence caused by the planar substrate, while non-radiative decay rate effects are negligible. We expect that this method to go beyond ensemble average quantities will find wider use to characterize a plethora of solid state emitters of current interest,  and can be further extended to also deal with, \emph{e.g.}, the statistics of blinking or  quantifying decay rates for bright and gray states separately.

\section{Materials and Methods}
We use a high tolerance ball lens from Edmund Optics, with a diameter of $4.00\unit{mm}$, a diameter tolerance of $+0.0/-3.00\unit{\mu m}$, and a maximal deviation from a sphere of $2\unit{\mu m}$. The ball lens is coated by physical vapor deposition of a $5\unit{nm}$ Ge adhesion layer, followed by $100\unit{nm}$ Ag. Finally, as a protection against scratches and oxidation, a   $35\unit{nm}$ layer  of $\mathrm{SiO_2}$ is evaporated onto the sphere. For easier handling and to prevent the ball lens from rolling over the sample,  we glued three identical ball lenses onto a cover glass slide in a tripod configuration prior to coating.  
A few considerations restrict the choice of radius of the spherical lens. In our measurement scheme, the distance to the mirror is increased by moving out from the contact point between the substrate and mirror. 
Firstly, in order for the configuration to best  approximate a  planar mirror for which the LDOS is easily calculated, we require that the radius is much larger than the emission wavelength. Secondly,  we require that the maximum difference in vertical distance  within the width of the signal collection area (in our case a diffraction-limited focus), is much smaller than the emission wavelength. Both of these requirements are easily fulfilled using macroscopic ball lenses with diameters of $\sim 1\unit{mm}$. 

Following previous work\cite{Carbone2007}, we synthesized CdSe/CdS dot-in-rod structures, \emph{i.e.} a spherical CdSe core embedded in a rod shaped CdS shell, for use as emitters. The CdSe core has a diameter of $3.2\unit{nm}$ and the shell has the dimensions $5.6\unit{nm}$ and $21\unit{nm}$ along the short and long axis, respectively. The nanorods (NRs) are covered by $15-20\unit{nm}$ of silica following ref. \cite{Koole2008}. The $\mathrm{SiO}_2$ coating imparts water solubility while providing an inert protection layer to reduce degradation of optical properties in aqueous environments. Moreover, for our experiment, the SiO$_2$ layer hinders electric coupling between clustered rods and reduces  aggregation when placing NRs on substrates. The NRs show an ensemble emission spectrum with a centre wavelength of $610 \unit{nm}$ and a full width at half maximum of $\sim 50\unit{nm}$. This width is largely due to size inhomogeneity, as single quantum dot-in-rod emitters of this type have spectral widths of around 10 nm (at room temperature, and at the pixel integration times used in this work), see supporting material available online. Since the silica coating makes the NRs hydrophilic, in order to achieve a homogeneously distributed single layer of NRs, we use a Piranha cleaned glass substrate followed by a 5 min bath in HCl ($38 \%$) to render the surface hydrophilic. 
The NRs are subsequently spin-coated onto the substrate (500 RPM for 30 s) from a diluted ethanol dispersion using carefully tuned spin parameters and dilution to achieve a single layered homogeneous distribution with a density of order 1--10~$\mu$m$^{-2}$.

We use a confocal fluorescence lifetime imaging microscope, described in earlier work\cite{Frimmer2011}, in which a piezo stage allows translational scanning relative to the laser focus of the sample substrate containing a dilute surface coverage with the emitters, plus mirror tripod. Thereby, at different lateral position  we address emitters at a different vertical separation to the mirror surface.
The excitation  source is a $532\unit{nm}$, linearly polarized pulsed laser, with a pulse duration $<10\unit{ps}$ and a repetition rate of $10\unit{MHz}$, that is focused to a diffraction limited spot using a 100$\times$ oil-immersion objective (NA=1.4). To avoid excessive blinking, creation of biexcitons and saturation of our emitters we use a low pump power of $\sim10\unit{nW}$. The collected fluorescence is focused onto a 20~$\mu$m silicon avalanche photodiode (IdQuantique ULN)  that is connected to a Becker and Hickl DPC230 timing card registering the arrival times of laser pulses and fluorescence photons with  $165\unit{ps}$ resolution. The sum of background and dark counts was measured to be $\sim 4\unit{counts/s}$. We  scan with a step size of around 300 nm, comparable to the diffraction limit, and with a scan speed (50 Hz pixel clock) that is a trade off between focus drift  and sufficient collection time per pixel to fit a lifetime to the detected signal.

\begin{acknowledgments}  
We are grateful to M. Frimmer and A. Mohtashami for help in the initial stages of the project. This work is part of the research program of the "Foundation for Fundamental Research on Matter (FOM)", which is financially supported by the "The Netherlands Organization for Scientific Research (NWO)". AFK gratefully acknowledges a NWO-Vidi grant for financial support. PL acknowledges support by the Carlsberg Foundation as well as the Danish Research Council for Independent Research (Grant no. FTP 11-116740).\\
\end{acknowledgments}

\emph{Supporting Information Available}: Measured spectra of single nanorods as well as ensemble; decay rate analysis of a single nanorod. This material is available free of charge via the Internet at \url{http://pubs.acs.org}.

\bibliography{Lunnemann_Arxiv}

\begin{thebibliography}{41}%
\makeatletter
\providecommand \@ifxundefined [1]{%
 \@ifx{#1\undefined}
}%
\providecommand \@ifnum [1]{%
 \ifnum #1\expandafter \@firstoftwo
 \else \expandafter \@secondoftwo
 \fi
}%
\providecommand \@ifx [1]{%
 \ifx #1\expandafter \@firstoftwo
 \else \expandafter \@secondoftwo
 \fi
}%
\providecommand \natexlab [1]{#1}%
\providecommand \enquote  [1]{``#1''}%
\providecommand \bibnamefont  [1]{#1}%
\providecommand \bibfnamefont [1]{#1}%
\providecommand \citenamefont [1]{#1}%
\providecommand \href@noop [0]{\@secondoftwo}%
\providecommand \href [0]{\begingroup \@sanitize@url \@href}%
\providecommand \@href[1]{\@@startlink{#1}\@@href}%
\providecommand \@@href[1]{\endgroup#1\@@endlink}%
\providecommand \@sanitize@url [0]{\catcode `\\12\catcode `\$12\catcode
  `\&12\catcode `\#12\catcode `\^12\catcode `\_12\catcode `\%12\relax}%
\providecommand \@@startlink[1]{}%
\providecommand \@@endlink[0]{}%
\providecommand \url  [0]{\begingroup\@sanitize@url \@url }%
\providecommand \@url [1]{\endgroup\@href {#1}{\urlprefix }}%
\providecommand \urlprefix  [0]{URL }%
\providecommand \Eprint [0]{\href }%
\providecommand \doibase [0]{http://dx.doi.org/}%
\providecommand \selectlanguage [0]{\@gobble}%
\providecommand \bibinfo  [0]{\@secondoftwo}%
\providecommand \bibfield  [0]{\@secondoftwo}%
\providecommand \translation [1]{[#1]}%
\providecommand \BibitemOpen [0]{}%
\providecommand \bibitemStop [0]{}%
\providecommand \bibitemNoStop [0]{.\EOS\space}%
\providecommand \EOS [0]{\spacefactor3000\relax}%
\providecommand \BibitemShut  [1]{\csname bibitem#1\endcsname}%
\let\auto@bib@innerbib\@empty
%</preamble>
\bibitem [{\citenamefont {Resch-Genger}\ \emph {et~al.}(2008)\citenamefont
  {Resch-Genger}, \citenamefont {Grabolle}, \citenamefont {Cavaliere-Jaricot},
  \citenamefont {Nitschke},\ and\ \citenamefont {Nann}}]{Resch-Genger2008}%
  \BibitemOpen
  \bibfield  {author} {\bibinfo {author} {\bibfnamefont {U.}~\bibnamefont
  {Resch-Genger}}, \bibinfo {author} {\bibfnamefont {M.}~\bibnamefont
  {Grabolle}}, \bibinfo {author} {\bibfnamefont {S.}~\bibnamefont
  {Cavaliere-Jaricot}}, \bibinfo {author} {\bibfnamefont {R.}~\bibnamefont
  {Nitschke}}, \ and\ \bibinfo {author} {\bibfnamefont {T.}~\bibnamefont
  {Nann}},\ }\href@noop {} {\bibfield  {journal} {\bibinfo  {journal} {Nat.
  Methods}\ }\textbf {\bibinfo {volume} {5}},\ \bibinfo {pages} {763} (\bibinfo
  {year} {2008})}\BibitemShut {NoStop}%
\bibitem [{\citenamefont {Lichtman}\ and\ \citenamefont
  {Conchello}(2005)}]{Lichtman2005}%
  \BibitemOpen
  \bibfield  {author} {\bibinfo {author} {\bibfnamefont {J.~W.}\ \bibnamefont
  {Lichtman}}\ and\ \bibinfo {author} {\bibfnamefont {J.-A.}\ \bibnamefont
  {Conchello}},\ }\href@noop {} {\bibfield  {journal} {\bibinfo  {journal}
  {Nat. Methods}\ }\textbf {\bibinfo {volume} {2}},\ \bibinfo {pages} {910}
  (\bibinfo {year} {2005})}\BibitemShut {NoStop}%
\bibitem [{\citenamefont {Caruge}\ \emph {et~al.}(2008)\citenamefont {Caruge},
  \citenamefont {Halpert}, \citenamefont {Wood}, \citenamefont {Bulovi\'{c}},\
  and\ \citenamefont {Bawendi}}]{Caruge2008}%
  \BibitemOpen
  \bibfield  {author} {\bibinfo {author} {\bibfnamefont {J.~M.}\ \bibnamefont
  {Caruge}}, \bibinfo {author} {\bibfnamefont {J.~E.}\ \bibnamefont {Halpert}},
  \bibinfo {author} {\bibfnamefont {V.}~\bibnamefont {Wood}}, \bibinfo {author}
  {\bibfnamefont {V.}~\bibnamefont {Bulovi\'{c}}}, \ and\ \bibinfo {author}
  {\bibfnamefont {M.~G.}\ \bibnamefont {Bawendi}},\ }\href@noop {} {\bibfield
  {journal} {\bibinfo  {journal} {Nat. Photonics}\ }\textbf {\bibinfo {volume}
  {2}},\ \bibinfo {pages} {247} (\bibinfo {year} {2008})}\BibitemShut {NoStop}%
\bibitem [{\citenamefont {{van Der Poel}}\ \emph {et~al.}(2006)\citenamefont
  {{van Der Poel}}, \citenamefont {M{\o}rk}, \citenamefont {Somers},
  \citenamefont {Forchel}, \citenamefont {Reithmaier},\ and\ \citenamefont
  {Eisenstein}}]{VanderPoel2006}%
  \BibitemOpen
  \bibfield  {author} {\bibinfo {author} {\bibfnamefont {M.}~\bibnamefont {{van
  Der Poel}}}, \bibinfo {author} {\bibfnamefont {J.}~\bibnamefont {M{\o}rk}},
  \bibinfo {author} {\bibfnamefont {A.}~\bibnamefont {Somers}}, \bibinfo
  {author} {\bibfnamefont {A.}~\bibnamefont {Forchel}}, \bibinfo {author}
  {\bibfnamefont {J.~P.}\ \bibnamefont {Reithmaier}}, \ and\ \bibinfo {author}
  {\bibfnamefont {G.}~\bibnamefont {Eisenstein}},\ }\href@noop {} {\bibfield
  {journal} {\bibinfo  {journal} {Appl. Phys. Lett.}\ }\textbf {\bibinfo
  {volume} {89}},\ \bibinfo {pages} {81102} (\bibinfo {year}
  {2006})}\BibitemShut {NoStop}%
\bibitem [{\citenamefont {Eliseev}\ \emph {et~al.}(2001)\citenamefont
  {Eliseev}, \citenamefont {Li}, \citenamefont {Liu}, \citenamefont {Newell},
  \citenamefont {Lester},\ and\ \citenamefont {Malloy}}]{Eliseev2001}%
  \BibitemOpen
  \bibfield  {author} {\bibinfo {author} {\bibfnamefont {P.}~\bibnamefont
  {Eliseev}}, \bibinfo {author} {\bibfnamefont {H.}~\bibnamefont {Li}},
  \bibinfo {author} {\bibfnamefont {T.}~\bibnamefont {Liu}}, \bibinfo {author}
  {\bibfnamefont {T.}~\bibnamefont {Newell}}, \bibinfo {author} {\bibfnamefont
  {L.}~\bibnamefont {Lester}}, \ and\ \bibinfo {author} {\bibfnamefont
  {K.}~\bibnamefont {Malloy}},\ }\href@noop {} {\bibfield  {journal} {\bibinfo
  {journal} {IEEE J. Sel. Top. Quant. Electron.}\ }\textbf {\bibinfo {volume}
  {7}},\ \bibinfo {pages} {135} (\bibinfo {year} {2001})}\BibitemShut {NoStop}%
\bibitem [{\citenamefont {Claudon}\ \emph {et~al.}(2010)\citenamefont
  {Claudon}, \citenamefont {Bleuse}, \citenamefont {Malik}, \citenamefont
  {Bazin}, \citenamefont {Jaffrennou}, \citenamefont {Gregersen}, \citenamefont
  {Sauvan}, \citenamefont {Lalanne},\ and\ \citenamefont
  {G\'{e}rard}}]{Malik2010}%
  \BibitemOpen
  \bibfield  {author} {\bibinfo {author} {\bibfnamefont {J.}~\bibnamefont
  {Claudon}}, \bibinfo {author} {\bibfnamefont {J.}~\bibnamefont {Bleuse}},
  \bibinfo {author} {\bibfnamefont {N.~S.}\ \bibnamefont {Malik}}, \bibinfo
  {author} {\bibfnamefont {M.}~\bibnamefont {Bazin}}, \bibinfo {author}
  {\bibfnamefont {P.}~\bibnamefont {Jaffrennou}}, \bibinfo {author}
  {\bibfnamefont {N.}~\bibnamefont {Gregersen}}, \bibinfo {author}
  {\bibfnamefont {C.}~\bibnamefont {Sauvan}}, \bibinfo {author} {\bibfnamefont
  {P.}~\bibnamefont {Lalanne}}, \ and\ \bibinfo {author} {\bibfnamefont
  {J.-M.}\ \bibnamefont {G\'{e}rard}},\ }\href@noop {} {\bibfield  {journal}
  {\bibinfo  {journal} {Nat. Photonics}\ }\textbf {\bibinfo {volume} {4}},\
  \bibinfo {pages} {174} (\bibinfo {year} {2010})}\BibitemShut {NoStop}%
\bibitem [{\citenamefont {Hennessy}\ \emph {et~al.}(2007)\citenamefont
  {Hennessy}, \citenamefont {Badolato}, \citenamefont {Winger}, \citenamefont
  {Gerace}, \citenamefont {Atat\"{u}re}, \citenamefont {Gulde}, \citenamefont
  {F\"{a}lt}, \citenamefont {Hu},\ and\ \citenamefont
  {Imamoglu}}]{Hennessy2007a}%
  \BibitemOpen
  \bibfield  {author} {\bibinfo {author} {\bibfnamefont {K.}~\bibnamefont
  {Hennessy}}, \bibinfo {author} {\bibfnamefont {a.}~\bibnamefont {Badolato}},
  \bibinfo {author} {\bibfnamefont {M.}~\bibnamefont {Winger}}, \bibinfo
  {author} {\bibfnamefont {D.}~\bibnamefont {Gerace}}, \bibinfo {author}
  {\bibfnamefont {M.}~\bibnamefont {Atat\"{u}re}}, \bibinfo {author}
  {\bibfnamefont {S.}~\bibnamefont {Gulde}}, \bibinfo {author} {\bibfnamefont
  {S.}~\bibnamefont {F\"{a}lt}}, \bibinfo {author} {\bibfnamefont {E.~L.}\
  \bibnamefont {Hu}}, \ and\ \bibinfo {author} {\bibfnamefont {A.}~\bibnamefont
  {Imamoglu}},\ }\href@noop {} {\bibfield  {journal} {\bibinfo  {journal}
  {Nature}\ }\textbf {\bibinfo {volume} {445}},\ \bibinfo {pages} {896}
  (\bibinfo {year} {2007})}\BibitemShut {NoStop}%
\bibitem [{\citenamefont {Lund-Hansen}\ \emph {et~al.}(2008)\citenamefont
  {Lund-Hansen}, \citenamefont {Stobbe}, \citenamefont {Julsgaard},
  \citenamefont {Thyrrestrup}, \citenamefont {S\"{u}nner}, \citenamefont
  {Kamp}, \citenamefont {Forchel},\ and\ \citenamefont
  {Lodahl}}]{Lund-Hansen2008}%
  \BibitemOpen
  \bibfield  {author} {\bibinfo {author} {\bibfnamefont {T.}~\bibnamefont
  {Lund-Hansen}}, \bibinfo {author} {\bibfnamefont {S.}~\bibnamefont {Stobbe}},
  \bibinfo {author} {\bibfnamefont {B.}~\bibnamefont {Julsgaard}}, \bibinfo
  {author} {\bibfnamefont {H.}~\bibnamefont {Thyrrestrup}}, \bibinfo {author}
  {\bibfnamefont {T.}~\bibnamefont {S\"{u}nner}}, \bibinfo {author}
  {\bibfnamefont {M.}~\bibnamefont {Kamp}}, \bibinfo {author} {\bibfnamefont
  {A.}~\bibnamefont {Forchel}}, \ and\ \bibinfo {author} {\bibfnamefont
  {P.}~\bibnamefont {Lodahl}},\ }\href@noop {} {\bibfield  {journal} {\bibinfo
  {journal} {Phys. Rev. Lett.}\ }\textbf {\bibinfo {volume} {101}},\ \bibinfo
  {pages} {1} (\bibinfo {year} {2008})}\BibitemShut {NoStop}%
\bibitem [{\citenamefont {Julsgaard}\ \emph {et~al.}(2008)\citenamefont
  {Julsgaard}, \citenamefont {Johansen}, \citenamefont {Stobbe}, \citenamefont
  {Stolberg-Rohr}, \citenamefont {S\"{u}nner}, \citenamefont {Kamp},
  \citenamefont {Forchel},\ and\ \citenamefont {Lodahl}}]{Julsgaard2008}%
  \BibitemOpen
  \bibfield  {author} {\bibinfo {author} {\bibfnamefont {B.}~\bibnamefont
  {Julsgaard}}, \bibinfo {author} {\bibfnamefont {J.}~\bibnamefont {Johansen}},
  \bibinfo {author} {\bibfnamefont {S.}~\bibnamefont {Stobbe}}, \bibinfo
  {author} {\bibfnamefont {T.}~\bibnamefont {Stolberg-Rohr}}, \bibinfo {author}
  {\bibfnamefont {T.}~\bibnamefont {S\"{u}nner}}, \bibinfo {author}
  {\bibfnamefont {M.}~\bibnamefont {Kamp}}, \bibinfo {author} {\bibfnamefont
  {A.}~\bibnamefont {Forchel}}, \ and\ \bibinfo {author} {\bibfnamefont
  {P.}~\bibnamefont {Lodahl}},\ }\href@noop {} {\bibfield  {journal} {\bibinfo
  {journal} {Appl. Phys. Lett.}\ }\textbf {\bibinfo {volume} {93}},\ \bibinfo
  {pages} {094102} (\bibinfo {year} {2008})}\BibitemShut {NoStop}%
\bibitem [{\citenamefont {Frimmer}\ \emph {et~al.}(2011)\citenamefont
  {Frimmer}, \citenamefont {Chen},\ and\ \citenamefont
  {Koenderink}}]{Frimmer2011}%
  \BibitemOpen
  \bibfield  {author} {\bibinfo {author} {\bibfnamefont {M.}~\bibnamefont
  {Frimmer}}, \bibinfo {author} {\bibfnamefont {Y.}~\bibnamefont {Chen}}, \
  and\ \bibinfo {author} {\bibfnamefont {A.~F.}\ \bibnamefont {Koenderink}},\
  }\href@noop {} {\bibfield  {journal} {\bibinfo  {journal} {Phys. Rev. Lett.}\
  }\textbf {\bibinfo {volume} {107}} (\bibinfo {year} {2011})}\BibitemShut
  {NoStop}%
\bibitem [{\citenamefont {Frimmer}\ and\ \citenamefont
  {Koenderink}(2012)}]{Frimmer2012}%
  \BibitemOpen
  \bibfield  {author} {\bibinfo {author} {\bibfnamefont {M.}~\bibnamefont
  {Frimmer}}\ and\ \bibinfo {author} {\bibfnamefont {A.~F.}\ \bibnamefont
  {Koenderink}},\ }\href@noop {} {\bibfield  {journal} {\bibinfo  {journal}
  {Phys. Rev. B}\ }\textbf {\bibinfo {volume} {86}},\ \bibinfo {pages} {235428}
  (\bibinfo {year} {2012})}\BibitemShut {NoStop}%
\bibitem [{\citenamefont {Sprik}\ \emph {et~al.}(1996)\citenamefont {Sprik},
  \citenamefont {van Tiggelen},\ and\ \citenamefont {Lagendijk}}]{Sprik1996}%
  \BibitemOpen
  \bibfield  {author} {\bibinfo {author} {\bibfnamefont {R.}~\bibnamefont
  {Sprik}}, \bibinfo {author} {\bibfnamefont {B.~A.}\ \bibnamefont {van
  Tiggelen}}, \ and\ \bibinfo {author} {\bibfnamefont {A.}~\bibnamefont
  {Lagendijk}},\ }\href@noop {} {\bibfield  {journal} {\bibinfo  {journal}
  {Eur. Phys. Lett.}\ }\textbf {\bibinfo {volume} {35}},\ \bibinfo {pages}
  {265} (\bibinfo {year} {1996})}\BibitemShut {NoStop}%
\bibitem [{\citenamefont {Loudon}(2000)}]{Loudon2000}%
  \BibitemOpen
  \bibfield  {author} {\bibinfo {author} {\bibfnamefont {R.}~\bibnamefont
  {Loudon}},\ }\href@noop {} {\emph {\bibinfo {title} {{The Quantum Theory of
  Light}}}},\ \bibinfo {edition} {3rd}\ ed.\ (\bibinfo  {publisher} {Oxford
  University Press},\ \bibinfo {address} {Oxford},\ \bibinfo {year} {2000})\
  p.\ \bibinfo {pages} {448}\BibitemShut {NoStop}%
\bibitem [{\citenamefont {Lodahl}\ \emph {et~al.}(2004)\citenamefont {Lodahl},
  \citenamefont {van Driel}, \citenamefont {Nikolaev}, \citenamefont {Irman},
  \citenamefont {Overgaag}, \citenamefont {Vanmaekelbergh},\ and\ \citenamefont
  {Vos}}]{Lodahl2004}%
  \BibitemOpen
  \bibfield  {author} {\bibinfo {author} {\bibfnamefont {P.}~\bibnamefont
  {Lodahl}}, \bibinfo {author} {\bibfnamefont {A.~.~F.}\ \bibnamefont {van
  Driel}}, \bibinfo {author} {\bibfnamefont {I.~S.}\ \bibnamefont {Nikolaev}},
  \bibinfo {author} {\bibfnamefont {A.}~\bibnamefont {Irman}}, \bibinfo
  {author} {\bibfnamefont {K.}~\bibnamefont {Overgaag}}, \bibinfo {author}
  {\bibfnamefont {D.}~\bibnamefont {Vanmaekelbergh}}, \ and\ \bibinfo {author}
  {\bibfnamefont {W.~L.}\ \bibnamefont {Vos}},\ }\href@noop {} {\bibfield
  {journal} {\bibinfo  {journal} {Nature}\ }\textbf {\bibinfo {volume} {430}},\
  \bibinfo {pages} {654} (\bibinfo {year} {2004})}\BibitemShut {NoStop}%
\bibitem [{\citenamefont {K\"{u}hn}\ \emph {et~al.}(2006)\citenamefont
  {K\"{u}hn}, \citenamefont {H{\aa}kanson}, \citenamefont {Rogobete},\ and\
  \citenamefont {Sandoghdar}}]{Kuhn2006}%
  \BibitemOpen
  \bibfield  {author} {\bibinfo {author} {\bibfnamefont {S.}~\bibnamefont
  {K\"{u}hn}}, \bibinfo {author} {\bibfnamefont {U.}~\bibnamefont
  {H{\aa}kanson}}, \bibinfo {author} {\bibfnamefont {L.}~\bibnamefont
  {Rogobete}}, \ and\ \bibinfo {author} {\bibfnamefont {V.}~\bibnamefont
  {Sandoghdar}},\ }\href@noop {} {\bibfield  {journal} {\bibinfo  {journal}
  {Phys. Rev. Lett.}\ }\textbf {\bibinfo {volume} {97}},\ \bibinfo {pages}
  {017402} (\bibinfo {year} {2006})}\BibitemShut {NoStop}%
\bibitem [{\citenamefont {Anger}\ \emph {et~al.}(2006)\citenamefont {Anger},
  \citenamefont {Bharadwaj},\ and\ \citenamefont {Novotny}}]{Anger2006}%
  \BibitemOpen
  \bibfield  {author} {\bibinfo {author} {\bibfnamefont {P.}~\bibnamefont
  {Anger}}, \bibinfo {author} {\bibfnamefont {P.}~\bibnamefont {Bharadwaj}}, \
  and\ \bibinfo {author} {\bibfnamefont {L.}~\bibnamefont {Novotny}},\
  }\href@noop {} {\bibfield  {journal} {\bibinfo  {journal} {Phys. Rev. Lett.}\
  }\textbf {\bibinfo {volume} {96}},\ \bibinfo {pages} {113002} (\bibinfo
  {year} {2006})}\BibitemShut {NoStop}%
\bibitem [{\citenamefont {Drexhage}\ \emph {et~al.}(1966)\citenamefont
  {Drexhage}, \citenamefont {Fleck}, \citenamefont {Sh\"{a}fer},\ and\
  \citenamefont {Sperling}}]{Drexhage1966}%
  \BibitemOpen
  \bibfield  {author} {\bibinfo {author} {\bibfnamefont {K.}~\bibnamefont
  {Drexhage}}, \bibinfo {author} {\bibfnamefont {M.}~\bibnamefont {Fleck}},
  \bibinfo {author} {\bibfnamefont {F.}~\bibnamefont {Sh\"{a}fer}}, \ and\
  \bibinfo {author} {\bibfnamefont {W.}~\bibnamefont {Sperling}},\ }\href@noop
  {} {\bibfield  {journal} {\bibinfo  {journal} {Ber. Bunsenges. Phys. Chem}\
  }\textbf {\bibinfo {volume} {20}},\ \bibinfo {pages} {1179} (\bibinfo {year}
  {1966})}\BibitemShut {NoStop}%
\bibitem [{\citenamefont {Snoeks}\ \emph {et~al.}(1995)\citenamefont {Snoeks},
  \citenamefont {Lagendijk},\ and\ \citenamefont {Polman}}]{Snoeks1995}%
  \BibitemOpen
  \bibfield  {author} {\bibinfo {author} {\bibfnamefont {E.}~\bibnamefont
  {Snoeks}}, \bibinfo {author} {\bibfnamefont {A.}~\bibnamefont {Lagendijk}}, \
  and\ \bibinfo {author} {\bibfnamefont {A.}~\bibnamefont {Polman}},\
  }\href@noop {} {\bibfield  {journal} {\bibinfo  {journal} {Phys. Rev. Lett.}\
  }\textbf {\bibinfo {volume} {74}},\ \bibinfo {pages} {2459} (\bibinfo {year}
  {1995})}\BibitemShut {NoStop}%
\bibitem [{\citenamefont {Amos}\ and\ \citenamefont {Barnes}(1997)}]{Amos1997}%
  \BibitemOpen
  \bibfield  {author} {\bibinfo {author} {\bibfnamefont {R.~M.}\ \bibnamefont
  {Amos}}\ and\ \bibinfo {author} {\bibfnamefont {W.~L.}\ \bibnamefont
  {Barnes}},\ }\href@noop {} {\bibfield  {journal} {\bibinfo  {journal} {Phys.
  Rev. B}\ }\textbf {\bibinfo {volume} {55}},\ \bibinfo {pages} {7249}
  (\bibinfo {year} {1997})}\BibitemShut {NoStop}%
\bibitem [{\citenamefont {Kwadrin}\ and\ \citenamefont
  {Koenderink}(2012)}]{Kwadrin2012}%
  \BibitemOpen
  \bibfield  {author} {\bibinfo {author} {\bibfnamefont {A.}~\bibnamefont
  {Kwadrin}}\ and\ \bibinfo {author} {\bibfnamefont {A.~F.}\ \bibnamefont
  {Koenderink}},\ }\href@noop {} {\bibfield  {journal} {\bibinfo  {journal} {J.
  Phys. Chem. C}\ }\textbf {\bibinfo {volume} {116}},\ \bibinfo {pages} {16666}
  (\bibinfo {year} {2012})}\BibitemShut {NoStop}%
\bibitem [{\citenamefont {Danz}\ \emph {et~al.}(2002)\citenamefont {Danz},
  \citenamefont {Heber}, \citenamefont {Br\"{a}uer},\ and\ \citenamefont
  {Kowarschik}}]{Danz2002}%
  \BibitemOpen
  \bibfield  {author} {\bibinfo {author} {\bibfnamefont {N.}~\bibnamefont
  {Danz}}, \bibinfo {author} {\bibfnamefont {J.}~\bibnamefont {Heber}},
  \bibinfo {author} {\bibfnamefont {A.}~\bibnamefont {Br\"{a}uer}}, \ and\
  \bibinfo {author} {\bibfnamefont {R.}~\bibnamefont {Kowarschik}},\
  }\href@noop {} {\bibfield  {journal} {\bibinfo  {journal} {Phys. Rev. A}\
  }\textbf {\bibinfo {volume} {66}},\ \bibinfo {pages} {063809} (\bibinfo
  {year} {2002})}\BibitemShut {NoStop}%
\bibitem [{\citenamefont {Johansen}\ \emph {et~al.}(2008)\citenamefont
  {Johansen}, \citenamefont {Stobbe}, \citenamefont {Nikolaev}, \citenamefont
  {Lund-Hansen}, \citenamefont {Kristensen}, \citenamefont {Hvam},
  \citenamefont {Vos},\ and\ \citenamefont {Lodahl}}]{Johansen2008}%
  \BibitemOpen
  \bibfield  {author} {\bibinfo {author} {\bibfnamefont {J.}~\bibnamefont
  {Johansen}}, \bibinfo {author} {\bibfnamefont {S.}~\bibnamefont {Stobbe}},
  \bibinfo {author} {\bibfnamefont {I.}~\bibnamefont {Nikolaev}}, \bibinfo
  {author} {\bibfnamefont {T.}~\bibnamefont {Lund-Hansen}}, \bibinfo {author}
  {\bibfnamefont {P.}~\bibnamefont {Kristensen}}, \bibinfo {author}
  {\bibfnamefont {J.~M.}\ \bibnamefont {Hvam}}, \bibinfo {author}
  {\bibfnamefont {W.}~\bibnamefont {Vos}}, \ and\ \bibinfo {author}
  {\bibfnamefont {P.}~\bibnamefont {Lodahl}},\ }\href@noop {} {\bibfield
  {journal} {\bibinfo  {journal} {Phys. Rev. B}\ }\textbf {\bibinfo {volume}
  {77}},\ \bibinfo {pages} {073303} (\bibinfo {year} {2008})}\BibitemShut
  {NoStop}%
\bibitem [{\citenamefont {Stobbe}\ \emph {et~al.}(2009)\citenamefont {Stobbe},
  \citenamefont {Johansen}, \citenamefont {Kristensen}, \citenamefont {Hvam},\
  and\ \citenamefont {Lodahl}}]{Stobbe2009}%
  \BibitemOpen
  \bibfield  {author} {\bibinfo {author} {\bibfnamefont {S.}~\bibnamefont
  {Stobbe}}, \bibinfo {author} {\bibfnamefont {J.}~\bibnamefont {Johansen}},
  \bibinfo {author} {\bibfnamefont {P.}~\bibnamefont {Kristensen}}, \bibinfo
  {author} {\bibfnamefont {J.~M.}\ \bibnamefont {Hvam}}, \ and\ \bibinfo
  {author} {\bibfnamefont {P.}~\bibnamefont {Lodahl}},\ }\href@noop {}
  {\bibfield  {journal} {\bibinfo  {journal} {Phys. Rev. B}\ }\textbf {\bibinfo
  {volume} {80}},\ \bibinfo {pages} {1} (\bibinfo {year} {2009})}\BibitemShut
  {NoStop}%
\bibitem [{\citenamefont {Andersen}\ \emph {et~al.}(2010)\citenamefont
  {Andersen}, \citenamefont {Stobbe}, \citenamefont {S{\o}rensen},\ and\
  \citenamefont {Lodahl}}]{Andersen2010}%
  \BibitemOpen
  \bibfield  {author} {\bibinfo {author} {\bibfnamefont {M.~L.}\ \bibnamefont
  {Andersen}}, \bibinfo {author} {\bibfnamefont {S.}~\bibnamefont {Stobbe}},
  \bibinfo {author} {\bibfnamefont {A.~S.}\ \bibnamefont {S{\o}rensen}}, \ and\
  \bibinfo {author} {\bibfnamefont {P.}~\bibnamefont {Lodahl}},\ }\href@noop {}
  {\bibfield  {journal} {\bibinfo  {journal} {Nat. Phys.}\ }\textbf {\bibinfo
  {volume} {7}},\ \bibinfo {pages} {215} (\bibinfo {year} {2010})}\BibitemShut
  {NoStop}%
\bibitem [{\citenamefont {Leistikow}\ \emph {et~al.}(2009)\citenamefont
  {Leistikow}, \citenamefont {Johansen}, \citenamefont {Kettelarij},
  \citenamefont {Lodahl},\ and\ \citenamefont {Vos}}]{Leistikow2009}%
  \BibitemOpen
  \bibfield  {author} {\bibinfo {author} {\bibfnamefont {M.}~\bibnamefont
  {Leistikow}}, \bibinfo {author} {\bibfnamefont {J.}~\bibnamefont {Johansen}},
  \bibinfo {author} {\bibfnamefont {A.}~\bibnamefont {Kettelarij}}, \bibinfo
  {author} {\bibfnamefont {P.}~\bibnamefont {Lodahl}}, \ and\ \bibinfo {author}
  {\bibfnamefont {W.}~\bibnamefont {Vos}},\ }\href@noop {} {\bibfield
  {journal} {\bibinfo  {journal} {Phys. Rev. B}\ }\textbf {\bibinfo {volume}
  {79}},\ \bibinfo {pages} {1} (\bibinfo {year} {2009})}\BibitemShut {NoStop}%
\bibitem [{\citenamefont {Brokmann}\ \emph {et~al.}(2004)\citenamefont
  {Brokmann}, \citenamefont {Coolen}, \citenamefont {Dahan},\ and\
  \citenamefont {Hermier}}]{Brokmann2004}%
  \BibitemOpen
  \bibfield  {author} {\bibinfo {author} {\bibfnamefont {X.}~\bibnamefont
  {Brokmann}}, \bibinfo {author} {\bibfnamefont {L.}~\bibnamefont {Coolen}},
  \bibinfo {author} {\bibfnamefont {M.}~\bibnamefont {Dahan}}, \ and\ \bibinfo
  {author} {\bibfnamefont {J.}~\bibnamefont {Hermier}},\ }\href@noop {}
  {\bibfield  {journal} {\bibinfo  {journal} {Phys. Rev. Lett.}\ }\textbf
  {\bibinfo {volume} {93}},\ \bibinfo {pages} {107403} (\bibinfo {year}
  {2004})}\BibitemShut {NoStop}%
\bibitem [{\citenamefont {Buchler}\ \emph {et~al.}(2005)\citenamefont
  {Buchler}, \citenamefont {Kalkbrenner}, \citenamefont {Hettich},\ and\
  \citenamefont {Sandoghdar}}]{Buchler2005}%
  \BibitemOpen
  \bibfield  {author} {\bibinfo {author} {\bibfnamefont {B.}~\bibnamefont
  {Buchler}}, \bibinfo {author} {\bibfnamefont {T.}~\bibnamefont
  {Kalkbrenner}}, \bibinfo {author} {\bibfnamefont {C.}~\bibnamefont
  {Hettich}}, \ and\ \bibinfo {author} {\bibfnamefont {V.}~\bibnamefont
  {Sandoghdar}},\ }\href@noop {} {\bibfield  {journal} {\bibinfo  {journal}
  {Phys. Rev. Lett.}\ }\textbf {\bibinfo {volume} {95}},\ \bibinfo {pages}
  {063003} (\bibinfo {year} {2005})}\BibitemShut {NoStop}%
\bibitem [{\citenamefont {Frimmer}\ \emph {et~al.}(2012)\citenamefont
  {Frimmer}, \citenamefont {Mohtashami},\ and\ \citenamefont
  {Koenderink}}]{Frimmer2012a}%
  \BibitemOpen
  \bibfield  {author} {\bibinfo {author} {\bibfnamefont {M.}~\bibnamefont
  {Frimmer}}, \bibinfo {author} {\bibfnamefont {A.}~\bibnamefont {Mohtashami}},
  \ and\ \bibinfo {author} {\bibfnamefont {A.~F.}\ \bibnamefont {Koenderink}},\
  }\href {http://arxiv.org/abs/1212.5081} {\bibfield  {journal} {\bibinfo
  {journal} {arXiv}\ ,\ \bibinfo {pages} {11}} (\bibinfo {year} {2012})},\
  \Eprint {http://arxiv.org/abs/1212.5081} {arXiv:1212.5081} \BibitemShut
  {NoStop}%
\bibitem [{\citenamefont {Chizhik}\ \emph {et~al.}(2011)\citenamefont
  {Chizhik}, \citenamefont {Chizhik}, \citenamefont {Khoptyar}, \citenamefont
  {B\"{ar}}, \citenamefont {Meixner},\ and\ \citenamefont
  {Enderlein}}]{Chizhik2011}%
  \BibitemOpen
  \bibfield  {author} {\bibinfo {author} {\bibfnamefont {A.~I.}\ \bibnamefont
  {Chizhik}}, \bibinfo {author} {\bibfnamefont {A.~M.}\ \bibnamefont
  {Chizhik}}, \bibinfo {author} {\bibfnamefont {D.}~\bibnamefont {Khoptyar}},
  \bibinfo {author} {\bibfnamefont {S.}~\bibnamefont {B\"{ar}}}, \bibinfo
  {author} {\bibfnamefont {A.~J.}\ \bibnamefont {Meixner}}, \ and\ \bibinfo
  {author} {\bibfnamefont {J.}~\bibnamefont {Enderlein}},\ }\href@noop {}
  {\bibfield  {journal} {\bibinfo  {journal} {Nano Lett.}\ }\textbf {\bibinfo
  {volume} {11}},\ \bibinfo {pages} {1700} (\bibinfo {year}
  {2011})}\BibitemShut {NoStop}%
\bibitem [{\citenamefont {Bajzer}\ \emph {et~al.}(1991)\citenamefont {Bajzer},
  \citenamefont {Therneau}, \citenamefont {Sharp},\ and\ \citenamefont
  {Prendergast}}]{Bajzer1991}%
  \BibitemOpen
  \bibfield  {author} {\bibinfo {author} {\bibfnamefont {Z.}~\bibnamefont
  {Bajzer}}, \bibinfo {author} {\bibfnamefont {T.~M.}\ \bibnamefont
  {Therneau}}, \bibinfo {author} {\bibfnamefont {J.~C.}\ \bibnamefont {Sharp}},
  \ and\ \bibinfo {author} {\bibfnamefont {F.~G.}\ \bibnamefont
  {Prendergast}},\ }\href@noop {} {\bibfield  {journal} {\bibinfo  {journal}
  {Eur. Biophys. J.}\ }\textbf {\bibinfo {volume} {20}},\ \bibinfo {pages}
  {247} (\bibinfo {year} {1991})}\BibitemShut {NoStop}%
\bibitem [{\citenamefont {Galland}\ \emph {et~al.}(2011)\citenamefont
  {Galland}, \citenamefont {Ghosh}, \citenamefont {Steinbr\"{u}ck},
  \citenamefont {Sykora}, \citenamefont {Hollingsworth}, \citenamefont
  {Klimov},\ and\ \citenamefont {Htoon}}]{Galland2011}%
  \BibitemOpen
  \bibfield  {author} {\bibinfo {author} {\bibfnamefont {C.}~\bibnamefont
  {Galland}}, \bibinfo {author} {\bibfnamefont {Y.}~\bibnamefont {Ghosh}},
  \bibinfo {author} {\bibfnamefont {A.}~\bibnamefont {Steinbr\"{u}ck}},
  \bibinfo {author} {\bibfnamefont {M.}~\bibnamefont {Sykora}}, \bibinfo
  {author} {\bibfnamefont {J.~a.}\ \bibnamefont {Hollingsworth}}, \bibinfo
  {author} {\bibfnamefont {V.~I.}\ \bibnamefont {Klimov}}, \ and\ \bibinfo
  {author} {\bibfnamefont {H.}~\bibnamefont {Htoon}},\ }\href@noop {}
  {\bibfield  {journal} {\bibinfo  {journal} {Nature}\ }\textbf {\bibinfo
  {volume} {479}},\ \bibinfo {pages} {203} (\bibinfo {year}
  {2011})}\BibitemShut {NoStop}%
\bibitem [{\citenamefont {Paulus}\ \emph {et~al.}(2000)\citenamefont {Paulus},
  \citenamefont {Gay-Balmaz},\ and\ \citenamefont {Martin}}]{Paulus2000}%
  \BibitemOpen
  \bibfield  {author} {\bibinfo {author} {\bibfnamefont {M.}~\bibnamefont
  {Paulus}}, \bibinfo {author} {\bibfnamefont {P.}~\bibnamefont {Gay-Balmaz}},
  \ and\ \bibinfo {author} {\bibfnamefont {O.~J.~F.}\ \bibnamefont {Martin}},\
  }\href@noop {} {\bibfield  {journal} {\bibinfo  {journal} {Phys. Rev. E}\
  }\textbf {\bibinfo {volume} {62}},\ \bibinfo {pages} {5797} (\bibinfo {year}
  {2000})}\BibitemShut {NoStop}%
\bibitem [{\citenamefont {Carbone}\ \emph {et~al.}(2007)\citenamefont
  {Carbone}, \citenamefont {Nobile}, \citenamefont {{De Giorgi}}, \citenamefont
  {Sala}, \citenamefont {Morello}, \citenamefont {Pompa}, \citenamefont
  {Hytch}, \citenamefont {Snoeck}, \citenamefont {Fiore}, \citenamefont
  {Franchini}, \citenamefont {Nadasan}, \citenamefont {Silvestre},
  \citenamefont {Chiodo}, \citenamefont {Kudera}, \citenamefont {Cingolani},
  \citenamefont {Krahne},\ and\ \citenamefont {Manna}}]{Carbone2007}%
  \BibitemOpen
  \bibfield  {author} {\bibinfo {author} {\bibfnamefont {L.}~\bibnamefont
  {Carbone}}, \bibinfo {author} {\bibfnamefont {C.}~\bibnamefont {Nobile}},
  \bibinfo {author} {\bibfnamefont {M.}~\bibnamefont {{De Giorgi}}}, \bibinfo
  {author} {\bibfnamefont {F.~D.}\ \bibnamefont {Sala}}, \bibinfo {author}
  {\bibfnamefont {G.}~\bibnamefont {Morello}}, \bibinfo {author} {\bibfnamefont
  {P.}~\bibnamefont {Pompa}}, \bibinfo {author} {\bibfnamefont
  {M.}~\bibnamefont {Hytch}}, \bibinfo {author} {\bibfnamefont
  {E.}~\bibnamefont {Snoeck}}, \bibinfo {author} {\bibfnamefont
  {A.}~\bibnamefont {Fiore}}, \bibinfo {author} {\bibfnamefont {I.~R.}\
  \bibnamefont {Franchini}}, \bibinfo {author} {\bibfnamefont {M.}~\bibnamefont
  {Nadasan}}, \bibinfo {author} {\bibfnamefont {A.~F.}\ \bibnamefont
  {Silvestre}}, \bibinfo {author} {\bibfnamefont {L.}~\bibnamefont {Chiodo}},
  \bibinfo {author} {\bibfnamefont {S.}~\bibnamefont {Kudera}}, \bibinfo
  {author} {\bibfnamefont {R.}~\bibnamefont {Cingolani}}, \bibinfo {author}
  {\bibfnamefont {R.}~\bibnamefont {Krahne}}, \ and\ \bibinfo {author}
  {\bibfnamefont {L.}~\bibnamefont {Manna}},\ }\href@noop {} {\bibfield
  {journal} {\bibinfo  {journal} {Nano Lett.}\ }\textbf {\bibinfo {volume}
  {7}},\ \bibinfo {pages} {2942} (\bibinfo {year} {2007})}\BibitemShut
  {NoStop}%
\bibitem [{\citenamefont {Talapin}\ \emph {et~al.}(2003)\citenamefont
  {Talapin}, \citenamefont {Koeppe}, \citenamefont {G\"{o}tzinger},
  \citenamefont {Kornowski}, \citenamefont {Lupton}, \citenamefont {Rogach},
  \citenamefont {Benson}, \citenamefont {Feldmann},\ and\ \citenamefont
  {Weller}}]{Talapin2003}%
  \BibitemOpen
  \bibfield  {author} {\bibinfo {author} {\bibfnamefont {D.~V.}\ \bibnamefont
  {Talapin}}, \bibinfo {author} {\bibfnamefont {R.}~\bibnamefont {Koeppe}},
  \bibinfo {author} {\bibfnamefont {S.}~\bibnamefont {G\"{o}tzinger}}, \bibinfo
  {author} {\bibfnamefont {A.}~\bibnamefont {Kornowski}}, \bibinfo {author}
  {\bibfnamefont {J.~M.}\ \bibnamefont {Lupton}}, \bibinfo {author}
  {\bibfnamefont {A.~L.}\ \bibnamefont {Rogach}}, \bibinfo {author}
  {\bibfnamefont {O.}~\bibnamefont {Benson}}, \bibinfo {author} {\bibfnamefont
  {J.}~\bibnamefont {Feldmann}}, \ and\ \bibinfo {author} {\bibfnamefont
  {H.}~\bibnamefont {Weller}},\ }\href@noop {} {\bibfield  {journal} {\bibinfo
  {journal} {Nano Lett.}\ }\textbf {\bibinfo {volume} {3}},\ \bibinfo {pages}
  {1677} (\bibinfo {year} {2003})}\BibitemShut {NoStop}%
\bibitem [{\citenamefont {van Driel}\ \emph {et~al.}(2007)\citenamefont {van
  Driel}, \citenamefont {Nikolaev}, \citenamefont {Vergeer}, \citenamefont
  {Lodahl}, \citenamefont {Vanmaekelbergh},\ and\ \citenamefont
  {Vos}}]{VanDriel2007}%
  \BibitemOpen
  \bibfield  {author} {\bibinfo {author} {\bibfnamefont {A.~F.}\ \bibnamefont
  {van Driel}}, \bibinfo {author} {\bibfnamefont {I.}~\bibnamefont {Nikolaev}},
  \bibinfo {author} {\bibfnamefont {P.}~\bibnamefont {Vergeer}}, \bibinfo
  {author} {\bibfnamefont {P.}~\bibnamefont {Lodahl}}, \bibinfo {author}
  {\bibfnamefont {D.}~\bibnamefont {Vanmaekelbergh}}, \ and\ \bibinfo {author}
  {\bibfnamefont {W.~L.}\ \bibnamefont {Vos}},\ }\href@noop {} {\bibfield
  {journal} {\bibinfo  {journal} {Phys. Rev. B}\ }\textbf {\bibinfo {volume}
  {75}},\ \bibinfo {pages} {035329} (\bibinfo {year} {2007})}\BibitemShut
  {NoStop}%
\bibitem [{\citenamefont {Nikolaev}\ \emph {et~al.}(2007)\citenamefont
  {Nikolaev}, \citenamefont {Lodahl}, \citenamefont {van Driel}, \citenamefont
  {Koenderink},\ and\ \citenamefont {Vos}}]{Nikolaev2007}%
  \BibitemOpen
  \bibfield  {author} {\bibinfo {author} {\bibfnamefont {I.}~\bibnamefont
  {Nikolaev}}, \bibinfo {author} {\bibfnamefont {P.}~\bibnamefont {Lodahl}},
  \bibinfo {author} {\bibfnamefont {A.~F.}\ \bibnamefont {van Driel}}, \bibinfo
  {author} {\bibfnamefont {A.~F.}\ \bibnamefont {Koenderink}}, \ and\ \bibinfo
  {author} {\bibfnamefont {W.~L.}\ \bibnamefont {Vos}},\ }\href@noop {}
  {\bibfield  {journal} {\bibinfo  {journal} {Phys. Rev. B}\ }\textbf {\bibinfo
  {volume} {75}},\ \bibinfo {pages} {115302} (\bibinfo {year}
  {2007})}\BibitemShut {NoStop}%
\bibitem [{\citenamefont {Vos}\ \emph {et~al.}(2009)\citenamefont {Vos},
  \citenamefont {Koenderink},\ and\ \citenamefont {Nikolaev}}]{Vos2009}%
  \BibitemOpen
  \bibfield  {author} {\bibinfo {author} {\bibfnamefont {W.~L.}\ \bibnamefont
  {Vos}}, \bibinfo {author} {\bibfnamefont {A.~F.}\ \bibnamefont {Koenderink}},
  \ and\ \bibinfo {author} {\bibfnamefont {I.~S.}\ \bibnamefont {Nikolaev}},\
  }\href@noop {} {\bibfield  {journal} {\bibinfo  {journal} {Phys. Rev. A}\
  }\textbf {\bibinfo {volume} {80}},\ \bibinfo {pages} {053802} (\bibinfo
  {year} {2009})}\BibitemShut {NoStop}%
\bibitem [{\citenamefont {Galland}\ \emph {et~al.}(2012)\citenamefont
  {Galland}, \citenamefont {Ghosh}, \citenamefont {Steinbr\"{u}ck},
  \citenamefont {Hollingsworth}, \citenamefont {Htoon},\ and\ \citenamefont
  {Klimov}}]{Galland2012}%
  \BibitemOpen
  \bibfield  {author} {\bibinfo {author} {\bibfnamefont {C.}~\bibnamefont
  {Galland}}, \bibinfo {author} {\bibfnamefont {Y.}~\bibnamefont {Ghosh}},
  \bibinfo {author} {\bibfnamefont {A.}~\bibnamefont {Steinbr\"{u}ck}},
  \bibinfo {author} {\bibfnamefont {J.~a.}\ \bibnamefont {Hollingsworth}},
  \bibinfo {author} {\bibfnamefont {H.}~\bibnamefont {Htoon}}, \ and\ \bibinfo
  {author} {\bibfnamefont {V.~I.}\ \bibnamefont {Klimov}},\ }\href@noop {}
  {\bibfield  {journal} {\bibinfo  {journal} {Nat. Commun.}\ }\textbf {\bibinfo
  {volume} {3}},\ \bibinfo {pages} {908} (\bibinfo {year} {2012})}\BibitemShut
  {NoStop}%
\bibitem [{\citenamefont {Spinicelli}\ \emph {et~al.}(2009)\citenamefont
  {Spinicelli}, \citenamefont {Buil}, \citenamefont {Qu\'{e}lin}, \citenamefont
  {Mahler}, \citenamefont {Dubertret},\ and\ \citenamefont
  {Hermier}}]{Spinicelli2009}%
  \BibitemOpen
  \bibfield  {author} {\bibinfo {author} {\bibfnamefont {P.}~\bibnamefont
  {Spinicelli}}, \bibinfo {author} {\bibfnamefont {S.}~\bibnamefont {Buil}},
  \bibinfo {author} {\bibfnamefont {X.}~\bibnamefont {Qu\'{e}lin}}, \bibinfo
  {author} {\bibfnamefont {B.}~\bibnamefont {Mahler}}, \bibinfo {author}
  {\bibfnamefont {B.}~\bibnamefont {Dubertret}}, \ and\ \bibinfo {author}
  {\bibfnamefont {J.-P.}\ \bibnamefont {Hermier}},\ }\href@noop {} {\bibfield
  {journal} {\bibinfo  {journal} {Phys. Rev. Lett.}\ }\textbf {\bibinfo
  {volume} {102}},\ \bibinfo {pages} {1} (\bibinfo {year} {2009})}\BibitemShut
  {NoStop}%
\bibitem [{\citenamefont {Cordones}\ \emph {et~al.}(2011)\citenamefont
  {Cordones}, \citenamefont {Bixby},\ and\ \citenamefont
  {Leone}}]{Cordones2011}%
  \BibitemOpen
  \bibfield  {author} {\bibinfo {author} {\bibfnamefont {A.~a.}\ \bibnamefont
  {Cordones}}, \bibinfo {author} {\bibfnamefont {T.~J.}\ \bibnamefont {Bixby}},
  \ and\ \bibinfo {author} {\bibfnamefont {S.~R.}\ \bibnamefont {Leone}},\
  }\href@noop {} {\bibfield  {journal} {\bibinfo  {journal} {J. Phys. Chem. C}\
  }\textbf {\bibinfo {volume} {115}},\ \bibinfo {pages} {6341} (\bibinfo {year}
  {2011})}\BibitemShut {NoStop}%
\bibitem [{\citenamefont {Koole}\ \emph {et~al.}(2008)\citenamefont {Koole},
  \citenamefont {van Schooneveld}, \citenamefont {Hilhorst}, \citenamefont {{de
  Mello Doneg\'{a}}}, \citenamefont {Hart}, \citenamefont {van Blaaderen},
  \citenamefont {Vanmaekelbergh},\ and\ \citenamefont {Meijerink}}]{Koole2008}%
  \BibitemOpen
  \bibfield  {author} {\bibinfo {author} {\bibfnamefont {R.}~\bibnamefont
  {Koole}}, \bibinfo {author} {\bibfnamefont {M.~M.}\ \bibnamefont {van
  Schooneveld}}, \bibinfo {author} {\bibfnamefont {J.}~\bibnamefont
  {Hilhorst}}, \bibinfo {author} {\bibfnamefont {C.}~\bibnamefont {{de Mello
  Doneg\'{a}}}}, \bibinfo {author} {\bibfnamefont {D.~C.~Ê.}\ \bibnamefont
  {Hart}}, \bibinfo {author} {\bibfnamefont {A.}~\bibnamefont {van Blaaderen}},
  \bibinfo {author} {\bibfnamefont {D.}~\bibnamefont {Vanmaekelbergh}}, \ and\
  \bibinfo {author} {\bibfnamefont {A.}~\bibnamefont {Meijerink}},\ }\href@noop
  {} {\bibfield  {journal} {\bibinfo  {journal} {Chem. Mater.}\ }\textbf
  {\bibinfo {volume} {20}},\ \bibinfo {pages} {2503} (\bibinfo {year}
  {2008})}\BibitemShut {NoStop}%
\end{thebibliography}%
\end{document}